# A gradual approach for maximising user conversion without compromising experience with high visual intensity website elements

Jarosław Jankowski[1]  Juho Hamari[2,3]  Jarosław Wątróbski[4]

[1] Faculty of Computer Science and Information Technology, West Pomeranian University of Technology, Szczecin, Poland

[2] Gamification Group, Faculty of Information Technology and Communications, Tampere University, Tampere, Finland

[3] Gamification Group, Faculty of Humanities, University of Turku, Turku, Finland

[4] Faculty of Economics and Management, University of Szczecin, Szczecin, Poland

Correspondence: jjankowski@wi.zut.edu.pl

**Abstract**

**Purpose:** The study develops and tests a method that can gradually find a sweet spot between user experience and visual intensity of website elements to maximise user conversion with minimal adverse effect.

**Approach:** In the first phase of the study, we develop the method. In the second stage, we test and evaluate the method via an empirical study; also, an experiment was conducted within web interface with the gradual intensity of visual elements.

**Findings:** The findings reveal that negative response grows faster than conversion when the visual intensity of the web interface is increased. However, a saturation point, where there is coexistence between maximum conversion and minimum negative response, can be found.

**Practical implications:** The findings imply that efforts to attract user attention should be pursued with increased caution and that a gradual approach presented in this study helps in finding a site-specific sweet-spot for a level of visual intensity by incrementally adjusting the elements of the interface and tracking the changes in user behaviour.

**Originality/Value:** Web marketing and advertising professionals often face the dilemma of determining the optimal level of visual intensity of interface element. Excessive use of marketing component and attention-grabbing visual elements can lead to an inferior user experience and consequent user churn due to growing intrusiveness. At the same time, too little visual intensity can fail to steer users. The present study provides a gradual approach which aids in finding a balance between user experience and visual intensity, maximising user conversion and thus providing a practical solution for the problem.

1. Introduction

The visual intensity of website elements is commonly used in online marketing to steer user behaviour. The overuse of techniques to attract user attention can ultimately lead to an inferior user experience and even user churn (Moe, 2006; Nielsen and Huber, 2009). However, marketing elements that are insufficiently visible will fail to generate beneficial interactions for the website operator due to the phenomenon of banner blindness, which was first identified by Benway and Lane (1998) and later confirmed by other researchers (Burke et al., 2005; Wong, 2001). With many dependencies affecting user experience, website developers and marketers have looked for ways to analyse the effects of interactivity on web user attitude (Teo et al., 2003). Research suggests that more control to web users within interactive content to minimise negative impact should be provided (Chatterjee, 2008). For effective marketing, it is important to determine the point at which conversion represented by the number of acquired interactions relative to the number of visitors is maximised while influence is minimised to reduce adverse impact (Zha and Wu, 2014). Dedicated elements of interface affecting user behaviour are designed to generate interactions (Haubl and Murray, 2001), while other strategies assess the interaction of users and product (Harbich and Hassenzahl, 2011). Such is typically achieved by using interactive objects such as banner ads, pop-ups, landing pages, recommending interfaces or other active elements within websites that use call-to-action messages. Currently, efforts to find a sweet-spot between the optimal levels of visual intensity have fallen short, with most studies based on breaking cognitive processes (Zha and Wu, 2014) and static approaches with repeated exposures of the same object (Moe, 2006), aggregated measures for positive and negative results with attempts to use fuzzy modelling of the levels of influence and sustainable marketing strategies (Jankowski et al., 2016).

Research questions arise regarding how to find the optimal level of visual intensity of web elements to attract user attention without negatively affecting user experience. What is the best way to keep visual intensity at the lowest possible level while acquiring acceptable results?

The motivation of the present study is to propose an approach for detecting the optimal level of visual intensity where interactions can be elicited without unnecessarily invading users with content based on higher intensity. Therefore, a gradual approach for adjusting the visual intensity of web interfaces towards maximised conversions and lower negative influence on user experience was developed. The main goal is searching dynamically for a balance between conversion within an interface and negative effects represented by the decline in user experience. Visual intensity, and effects of compelling stimuli are explored to influence target users' arousal state. Elements with high intensity can negatively affect web users' experience, and low-intensity elements do not affect arousal state, which leads to lower influence on user behaviour. While it is difficult to adjust the intensity to optimal level a priori, the goal of the proposed gradual adjustment is searching for appropriate levels for each user or audience. The problem was investigated from two perspectives: first, by increasing the level of visual intensity of interactive content and second, by decreasing it. This gradual approach makes it possible to search for the saturation point beyond which increased intensity fails to increase the number of interactions.

This approach extends earlier studies by assigning visual intensity levels to elements to scale influence and monitor positive and negative feedback. Previous studies explored the problem of overcoming low click-throughs by using constant and varying banners featuring call-to-action messages with different texts and flashing elements (Chatterjee, 2005). Differences were not scaled between exposures, and only clicks were monitored, without taking into account negative effects. Other research based on sequential exposition of animations (Baccot et al., 2009) reported a wear-out effect for repeated static and animated content without content changes between repetitions (Lee et al., 2015). Previous studies have also revealed that during repeated exposures, users may ignore advertising content due to a learned enduring disposition (Sun et al., 2008) or repetition blindness (Mancero et al., 2007).

In contrast to previous studies that have altered the type of stimulus to overcome the habituation effect (Hsieh et al., 2012), the present study explores how changes in visual intensity affect the habituation effect and identifies the saturation point. The advantage of the present approach it the possibility to observe the growth of response and find the best-converting version to impact business goals and user experience positively. Experiments were conducted on real audiences, not in laboratory setting like in most of earlier studies.

An experiment within a real online environment where user behaviour was measured to investigate optimal levels of visual intensity of used elements was conducted. The approach was verified by using an experimental object with different levels of visual intensity attracting user attention and motivating users to perform interactions in the form of clicks. Elements used were different colours with low- and high-contrast background, low and high-intensity flashing effect, and verbal call-to-action texts with scaled influence. The intensity of visual elements was scaled from low levels with neutral colours, through medium levels using more contrasting colours, towards highest intensity with flashing effects.

While experimental verification was based on selected characteristics of visual elements, the proposed approach could be used with scaled sizes, sound effects with varying intensities, and other techniques to attract user attention with the use of visual elements.

The paper is structured as follows. Section 2 provides a literature review in the area of influence on web user behaviour. Section 3 presents the approach for incrementally adjusting the level of visual influence with the empirical study in Section 4. Results of experiments are presented in Section 5 and they are discussed in Section 6. Section 7 outlines conclusions and provides a summary of the presented research.

## 2. Literature review

After delivering visitors to a website from sources such as paid advertising and organic searches, the next stage of marketing efforts is usually converting visitors into customers. Various developmental areas of web design are targeted at building successful websites that encourage impulse buying (Shen and Khalifa, 2012). Relevant studies have analysed the conative reactions of web users to the dominant hue of a website and its influence on mental imagery (Khrouf and Frikha, 2016) as well as the role of web quality in consumer attitudes towards online shopping (Al-Debei et al., 2015). Website operators benefit from traffic obtained through clicks, signups, purchases, or downloads (Tarafdar and Zhang, 2005) and sales of virtual items and currencies (Hamari and Lehdonvirta, 2010; Hamari, 2015; Hamari and Keronen, 2017). Many factors influence the behaviour of online consumers on websites (Constantinides, 2004) and online games (Vashisht and Sreejesh, 2017). Loyalty programs (Meyer-Waarden, 2008), gamification factors (Hamari, 2017; Hamari and Koivisto, 2015), communication through persuasive technologies (Fogg, 2002; Hamari et al., 2014; Jawdat et al. 2011), intra-site banners (Chatterjee, 2005), proper localisation of advertisements within editorial content (Micu and Pentina, 2015), recommendation systems (Nanou et al., 2010), and customer reviews and ratings (Cheung et al., 2003; Kumar and Benbasat, 2006) are commonly used to increase the conversion rate.

**2.1 Attracting user attention with increased visual intensity**

Communication with web users is usually performed through visual or textual messages within the website. The primary goals of textual messages and graphical elements are to motivate the user to perform the desired actions (Zhou, 2004) and to create positive brand attitudes (Micu and Pentina, 2015). In addition to static content, animations may be used, and their intensity affects user attention and interactions (Hamborg et al., 2012). The visual intensity of message elements influences the arousal state of the target users, their attitude toward the message (Yoon et al., 1999), and the message's cognitive effects (Hansen and Krygowski, 1994). More compelling visual elements increase arousal (Reeves et al., 1999), but intensive messages can also break primary cognitive processes; for example, disturbing news can negatively affect other messages (Mundorf et al., 1990). Moreover, messages based on compelling stimuli result in slower reaction times (Lang et al., 1995), confirming that arousing messages require greater processing resources (Grabe et al., 2000). Thus, the effects of marketing content intensity create design dilemmas, and it is difficult to adjust the intensity to an optimal level a priori.

**2.2 Repeated content exposures and reduced user attention**

When exploring a website, users are repeatedly exposed to content such as advertising banners based on computational inclusion (Aznar et al., 2011) or elements of recommending interfaces (Nanou et al., 2010). Previous studies of advertising content have shown that repeated exposure can improve brand recognition, even when no clicks or comparable interactions are registered (Briggs and Hollis, 1997). However, click-through usually declines after the first exposure and drops to an even lower level after the third exposure (Chatterjee et al., 2003). The most important contact with marketing content is the first one; the click response to repeated exposure to passive ads within the same visit will be negative. Chatterjee (2005) extended research and examined the role of repetition in banner exposure in terms of the direct response, represented by the click rate, and the indirect response, which was based on memory and aided and unaided recall. The stimulant materials were based on graphical banners, with four variants of advertisements assigned randomly to each loaded website; the banners featured a call-to-action message using different texts and flashing elements. Exploratory and goal-oriented usage patterns were explored, with a focus on differences between repeated content (Chatterjee, 2005). Another study investigated the sequential exposition of advertising content within editorial content, including differences in the order and time of placement of static, video, and animated content (Baccot et al., 2009). These studies have shown that the overall effectiveness of marketing content increases with repetition, although a habituation (Portnoy and Marchionini, 2010) and wear-out effects are observed, especially for static content (Lee et al., 2015).

**2.3 Overcoming banner blindness with animations and flashing effects**

Experienced web users focus on finding information or completing online tasks. Their attention is only slightly attracted by graphical elements such as advertising banners or other components used for marketing purposes. These users ignore irrelevant content due to banner blindness (Benway and Lane, 1998) and a limited ability to process information (Lang, 2000). One strategy to overcome banner blindness is the inclusion of visual components with vivid effects and luminance (Turatto and Galfano, 2000) and flashing elements in addition to static content. Flashing elements direct attention to a specific section of the website, reducing the need to scan the whole screen (Hong et al., 2004). Other possibilities include animations with one-time or continuously animated content (Hamborg et al., 2012). Increased animation intensity has been shown to increase the frequency but not the duration of fixation on banners. Another strategy is the identification of entry points within the website and salient areas for attracting the user's visual attention (Masciocchi and Still, 2013). Alternatively, dishabituation seeks to overcome habituation by altering the type, magnitude or schedule of a stimulus within a website (Hsieh et al., 2012).

**2.4 Intrusive marketing content and declining user experience**

In general, marketing content delivery can be treated as a trade-off between the website operator's business model and the user's need to search content on the website (Krishnamurthy and Wills, 2006).

Zhang (2006) highlighted this conflict between providing content and marketing. Marketing techniques can lead to increased website intrusiveness (Li et al., 2002) and a negative response from users (Leggatt, 2008). Content providers need to understand the potential negative effects of ads on viewing performance and use marketing content with minimum distracting effects. Although intrusive elements can temporarily increase the click-through rate and improve recall compared to animations, they tend to negatively affect brand perception (Shrestha, 2006). Overuse of intrusive components leads to content avoidance behaviours and increased interest in applications that block advertising content (McCoy et al., 2007). Moreover, intrusive content makes searching for information difficult and increases users' cognitive load, inducing frustration and other negative emotions (Brajnik and Gabrielli, 2010). The resultant negative attitude towards the website may reduce the duration of a user's visit (Moe, 2006). Accordingly, websites employing extensive advertising space may suffer from declining audiences and negative feedback.

**2.5 Overview of methodologies**

Recent experimental research related to marketing content within digital environments has explored several different methodologies. Some laboratory experiments have targeted selected groups of participants with assigned tasks and questionnaires (Chatterjee, 2008; Zha and Wu, 2014; Zhang, 2006; Baccot et al., 2009). Other field experiments have employed working websites and analysed the behaviour of real audiences (Moe, 2006; Krishnamurthy and Wills, 2006). The tasks assigned to participants usually take the form of search tasks within a synthetic environment (Zhang, 2006; Benway and Lane, 1998) or natural user tasks within websites (Moe, 2006). The stimuli used in these experiments are based on real advertisements (Zha and Wu, 2014) as well as synthetic marketing content (Moe, 2006, Baccot et al., 2009; Chatterjee, 2005). In addition, in most studies, the experimental conditions perceived by the target groups differ from those of the control group. Factorial experimental plans have been used to compare intrusive advertising techniques with standard unchanged content (Zha and Wu, 2014; McCoy et al., 2007), to study user behaviour with and without pop-up windows (Moe, 2006), to detect attitudes and number of interactions upon advertising content exposure (Yoon et al., 1999), to analyse search tasks when various advertising content is included (Burke et al. 2005), and to compare advertisements in terms of interactivity, shape and animation techniques (Hong et al., 2004). If the goal of the research is to acquire behavioural patterns of natural web users, the sampling criteria should ensure that the survey participants do not include media-savvy subjects (Zha and Wu, 2014).

## 3. Assumptions and conceptual framework for the gradual approach

### 3.1 General assumptions

Most websites use call-to-action messages within interfaces to motivate users to perform the desired interactions, whether for external parties or for the website's own purposes. Previous studies of content repetition have mainly focused on online advertising and marketing content but have failed to distinguish differences in stimulus level between repetitions. By seeking a compromise between negative responses and outcomes in the form of click-throughs, this research extends available approaches for measuring responses based on changes in intensity levels between repetitions.

The gradual approach to changes in content adopted in this study can reduce banner blindness and repetition blindness. Changes in visual intensity increase content salience and overcome the wear-out effect. The method presented here adjusts saliency parameters based on the individual characteristics of the user. This gradual approach uses typical patterns based on sequential requests, with the ability to show different marketing content at each request. Accordingly, the appropriate level of visual intensity can be determined for each user or user segment.

Although showing low-intensity elements would result in a low conversion rate, this study assumes that the ability to influence user behaviour using elements with high visual intensity is limited. If the visual intensity level is increased beyond the saturation point, negative responses will occur instead of a positive effect, i.e., conversions. This negative response is due to users' limited ability to process information. The complexity of the problem emphasises the need for detailed research on the influence of different website elements on conversion, user experience, and compromise solutions.

**3.2 Conceptual framework**

In this section, a generalised view of the structure of web interface, defined herein as a decomposable interactive object is proposed. The presented framework makes it possible to analyse specific interface elements, the impact of their visual intensity on consumer behaviour and helps determine the relationship between level of visual intensity and the results. This approach is oriented towards compromise solutions using changing levels of visual intensity of components in communication; it concentrates on interactive elements with the main goal of motivating users to perform actions such as clicks or signups. The research focuses on user interface elements with call-to-action purposes and the concept of decomposition of interactive objects. Additionally, the research investigates parameters divided into quantitative and qualitative groups and elements with high visual intensity flashing with assigned frequency. Elements of a web interface with call-to-action functions can be identified as interactive objects integrating components that influence users and motivate them to perform interactions. For each object, a set of available components is determined by $E = \{E_1, E_2, ..., E_n\}$. For every $E_i$ there is a number of available variants $E_i = \{e_{i,1}, e_{i,2}, ..., e_{i,cnt(i)}\}$ where $cnt(i)$ describes the number of variants available for the $i$-th element. For each component $e_{i,j}$ the level of visual intensity $l_{i,j}$ can be assigned. As a result, a cumulative influence effect of all used elements on user attention is possible. We assume that total visual intensity can be represented by aggregated measure $AI_i$ of object $E_i$ consisting of $k$ elements, as in the following formula:

$$AI_i = \sum_{j=1}^{k}(l_{i,j} * w_i) \tag{1}$$

where $l_{i,j}$ is the intensity level defined during the design process, and $w_j$ is the rank assumed for a given element based on effect size, which defines visual intensity in relation to other elements.

The main purpose of the proposed approach is to maximise factors related to results acquired by website operators, represented by the positive response $R^+$, and to minimise the negative response $R^-$, measured by users' attempts to disable the interface. Measurements for such actions are positive response factors $R^+F$, which represent user conversion and determine the relation of a number of desired interactions to the number of website users $U_t$ in a given time $t$. The negative response factor $R^-F$ is computed similarly. In the case of multi-dimensional monitoring and realisation of advertising campaigns, one can distinguish the conversion for the chosen types of interaction and determined audiences. Conversion factor $R^+F_{L,i,t}$ for positive responses can be determined in relation to levels of interface influence $L=1,...,k$ and interaction type $i=1,...,m$ in the period of time $t$ according to the following formula:

$$R^+F_{L,i,t} = \frac{R^+_{L,i,t}}{U_{L,t}} \qquad (2)$$

where $R^+_{L,i,t}$ represents the number of interactions of the *i-th* type generated with an interface with visual intensity level $L$ in the period of time $t$. $U_{L,t}$ is the total number of website users in time $t$ with contact with an interface with intensity level $L$. Aside from conversion rates, negative impact on users represented by negative response $R^-F$ is measured and the number of negative interactions is computed according to formula:

$$R^-F_{L,i,t} = \frac{R^-_{L,i,t}}{U_{L,t}} \qquad (3)$$

where $R^-_{L,i,t}$ is the number of negative interactions of $i$ type, generated within a website with the interface at the level of visual intensity $L$ in the period of time $t$, $U_{L,t}$ is the total number of website users in time $t$. With the determination of effectiveness and measurement factors, the primary purpose of the optimisation procedure is to maximise the realisation of determined positive actions represented by $RF^+$ and minimise $RF^-$. When generating a version of the interactive component, selection is made from the elements of the given set, based on the selection function $SF_t(G,C)->E_L(G,C)$. It is responsible for selection of elements for users from group *G* for contact number *C,* intended to maximise positive and minimise negative effects resulting from aggregated visual influence *L*. The tests conducted aim to maximise the function of evaluating effects $R^+F=f(E_1, E_2, ..., E_k)$ and minimise $R^-F=f(E_1, E_2, ..., E_k)$.

**3.3 Example process**

The example process is shown in Fig. 1 where users are divided into groups. For each group, a different strategy is used when showing the web interface. For users in group $G_1$, with each webpage reloading, the same level of visual intensity of each element is used. With the first impression, influence values are applied at the lowest level {1,1,1}, and the level is maintained for other contacts. For users in group $G_2$, visual intensity is increased in each contact with the website. The first contact with the interface is based on a low-level {1,1,1}. Then step-by-step visual intensity is increased to {1,1,2}, {1,2,2}, {2,2,2}, {2,2,3}, {2,3,3}, {3,3,3}. Users in group $G_3$ start interaction with an interface with high visual intensity levels {3,3,3}, and the intensity level is decreased with each reload.

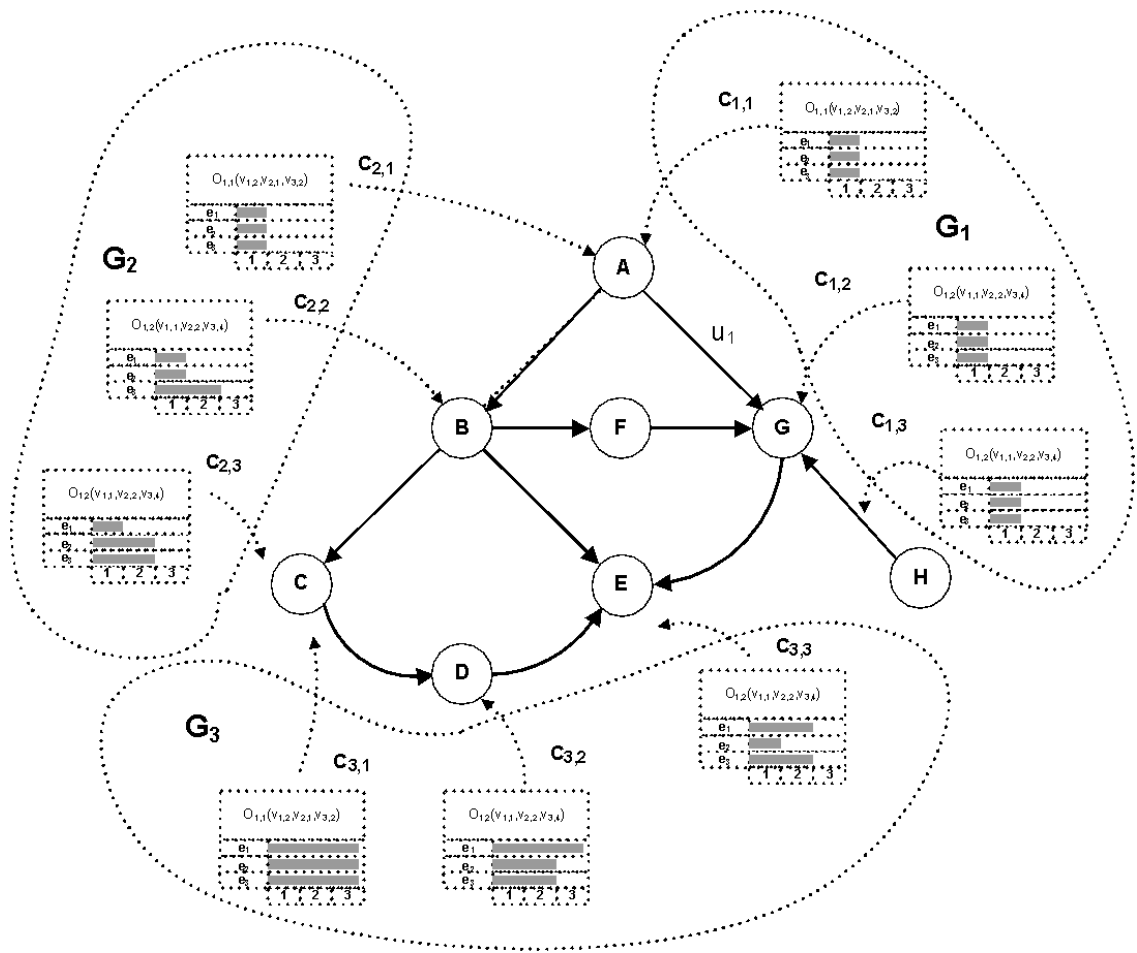

**Fig. 1.** Example process of interface changes during web browsing

The proposed approach concentrates on changing content and levels of visual intensity between repetitions and communication with content that can be changed over time when visiting sub-sites of a web page. Changes of states between *t* and *t+1* points of time affect user actions, and due to changes in stimuli, they can be more effective than repetitions of the same unchanged content. Presented research focused on an approach where the increase of levels is optimised so it avoids negative response from users. A framework is proposed to show how increasing and decreasing visual intensity levels between phases of communication affects the number of interactions. Large changes in levels can affect user experience and move attention from editorial content to marketing elements. At moderate levels, it can have a positive effect on the results of marketing actions, but excessive changes can result in negative side effects. The research showed how to detect patterns of user behaviours after changing object states.

**3.4 Sequential data processing**

Acquired data will be in the form of sequences; aggregating them will show results for each level of intensity. For sequence analysis and comparing positive and negative response, the dynamic time warping method can be used, targeted to various applications related to time series processing (Rabiner and Juang, 1993). The research method focuses on measuring the distance between sequence $R^+F$ and $R^-F$ and can be introduced in the form of similarity $S(R^+F, R^-F)$, which can be determined with use of available methods. In the easiest approach, the correlation factor can be utilised. The measure of similarity can be a distance among the Minkowski series as well (Niennattrakul et al., 2007). This is determined consistently with the following formula:

$$d(R^+F, R^-F) = \left( \sum_{i=1}^{n} \left| R^+F_i, R^-F_i \right|^P \right)^{1/P} \qquad (4)$$

In a case where P=2 is a distance among a series, it is a Euclidean distance. Determination of distance in this manner can cause limitations in cases with different length and a series with time-lag with changeable amplitude. The time warping distance method was applied to evaluate similarity. For two series $R^+F$ $(L_1, L_2, ..., L_n)$ and $R^-F(L_1, L_2, ..., L_m)$, with lengths respectively $n$, $m$ and $M$ matrix defined as interposing relations between series $R^+F$ and $R^-F$, $M_{i,j}$ element indicates the distance $d(x_i, y_j)$ between $x_i$ and $y_j$. Dependency among a series is determined by time warping path. The algorithm determines warping path with the lowest cost between two series. A series with a higher level of similarity can be easily compared because of alignment and dependencies caused by dynamic time distance. In the proposed procedure, the sequences can be compared to distinguish similarities and distance between $R^+F$ and $R^-F$ factors.

**3.5 Assumptions for integration with an online system**

Communicating with web users and motivating them to perform as desired through website operator actions is usually connected with sequences of contacts with website content in a traversal path. Messages are sent to a user during website content navigation. The proposed approach is based on the integration of a decision support system (DSS) with the online environment. While most approaches related to content servers, advertising servers or optimisation platforms are targeted mainly to maximise the number of interactions; the proposed approach is based on measuring positive and negative response. Fig. 2 presents the process of content selection within a website after receiving request $RQ_{i,j,k}$ from webpage $i$, user $j$ based on contact number $k$. A request is redirected to the content server (CS), and the selection from content database (CDB) is processed. This is based on parameters and selection function adjusting content and level of visual intensity to the parameters received with the request.

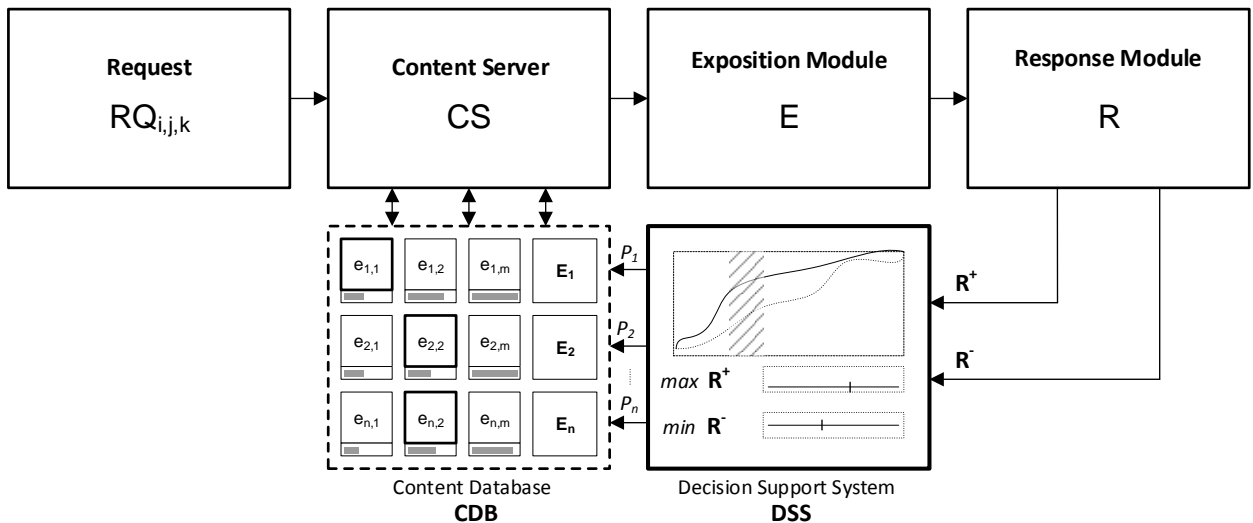

**Fig. 2.** Integration of systems based on adjustable levels of visual intensity with the content server

After selection, content is served and presented to the user with the exposition module $E$. Response module $R$ monitors positive $R^+$ and negative $R^-$ response and sends them as an input to DSS. Within the system; responses are aggregated, and results are compared. Balanced solutions based on satisfactory results with limited negative impact on the user are obtained and transferred in the form of parameters $P_1, P_2, \ldots, P_n$ to the content server are used for processing requests.

## 4. The empirical study design

When investigating human behaviour in information systems, experiments that manipulate information systems and gather behavioural data from real users can be considered internally valid. Positive and negative behavioural reactions are assumed to be due to visual intensity of presented experimental objects. In addition, compared to categorical experiments involving only the polar ends of visual intensity, the gradual approach increases the resolution of the results, thus enabling more accurate mapping of the point at which visual intensity reduces positive impressions. In the first stage, the experimental interface was designed following the mechanisms presented within the conceptual framework and integrated with the website. Users entering the website were assigned to five groups. For the first three groups, results were obtained for elements with stable and unchangeable levels of low, medium, or high visual intensity. Users in group five were targeted with increasing visual intensity with each page view in sequence and users in group four were receiving content with decreasing intensity. Following this procedure, the incremental approach was tested and compared with the flat and decreasing approaches.

### 4.1 Decomposition of the interface used in the experiment

The application was designed to be a 970 x 160 pixels-size interface within a website that contained casual games. Within the object, seven games from the same category like currently played game were shown and proposed to play. Fig. 3 presents a schematic structure of the object. Selected elements of the object had variants of visual intensity and were more attention-catching. The interface was dynamically generated each time the user selected a game from the catalogue. During the object's generation, seven games were randomly chosen from a database and were shown in the form of thumbnail images $I_1$-$I_7$, each 105 x 89 pixels; titles $T_1$-$T_7$ were shown above each of the thumbnails. One of the games was always shown as a featured game on the left side with an expanded description $D_1$, recommendation $R_1$, rating info $R_2$, and active background $BG_1$. Playing an advertised game was possible by clicking on a thumbnail or on button $B_1$ with call-to-action text $BT_1$. A header bar with the title $HT_1$ and background $BG_3$ invited users to play the recommended games. The users had the option to remove the recommending object after clicking button $C_1$ in the form of a, "[x]" sign that is well-known to users. Within the object, a mini-questionnaire was integrated asking the users' opinion about each form of recommending interface. Question $Q_1$ asked users whether the recommending interface was useful and should stay on the website. Possible answers were $A_1$-$A_7$, with $A_1$ for *strongly yes* up to $A_7$ for *strongly not*.

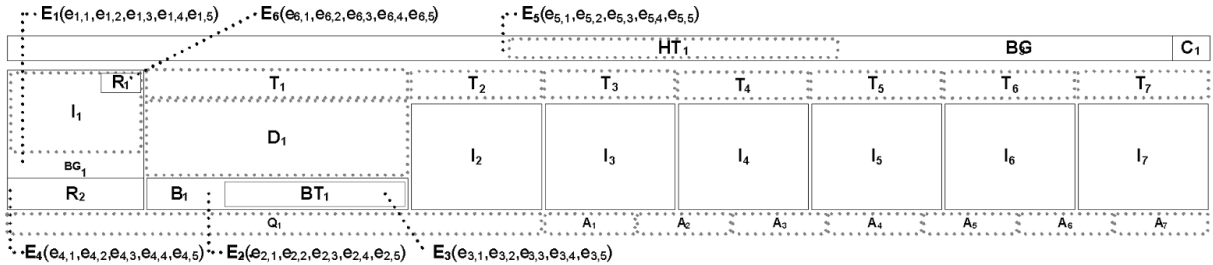

**Fig. 3.** Structure of experimental interface with thumbnail images $I_1$-$I_7$, titles $T_1$-$T_7$ above each of the thumbnails, an expanded description $D_1$, recommendation $R_1$, rating info $R_2$, active background $BG_1$, button $B_1$ with call-to-action text $BT_1$, a header bar with the title $HT_1$ and background $BG_3$ inviting users to play the recommended games, button $C_1$ for disabling recommending interface and question $Q_1$ with possible answers $A_1$-$A_7$.

Changeable elements were assigned to $E_1$-$E_6$ sets of available design variants with different visual intensity and influence on a user. Each component of the changeable set had a different purpose and influence on users' attention. Element $E_1$ is the border of a thumbnail image of the featured game, which can grab the user's attention by applying different visual intensities at five levels starting from blue at level one, trough more intensive orange at second and red at third level. The fourth stage was switching red with orange with low frequency 1/2 second for each colour while at the fifth level, highest frequency 1/12 second per each colour was used with highest ability to attract attention. Element $E_2$ represents button with five intensities similar to levels integrated with element $E_1$. Element $E_3$ contains call-to-action text with five visual intensities. Element $E_4$ is an element showing the rating for a selected game. At the first level, the element is empty, and no rating is shown. At levels two and three, a rating is shown in the form of static image stars from one star to three stars. When it reaches level four and five, the rating stars are flashing with low frequency and two changes per second up to high frequency and six changes per second, respectively. Element $E_5$ is a graphical header with the main goal of attracting user attention using five levels of visual intensity by contrasting and flashing effects. Element $E_6$ is a graphical bar with text "Recommended" shown in the layer above thumbnail with various colour levels and flashing effects.

The whole website is displayed with light-blue shades and elements on the website are blue tones. The header and elements of EC2 use different shades of blue. Level one of element $E_5$ marked by $e_{5,1}$ represents the colour of the bar and static image with low impact on the user by not contrasting with other parts of the website. Level two with an orange element $e_{5,2}$ contrasting with background and surrounding elements, red element $e_{5,3}$, $e_{5,4}$ flashing with two-coloured animations. Flickering effect was set to 12 frames per second, and each colour was shown during three frames, meaning there were two changes per second. The fifth level $e_{5,5}$ used similar timing for animation but with colours that were more contrasting to the background. The colours contrasted each other more than in level four and as a result due to higher visual intensity it had a greater interference on website content processing. Object states were changed between impressions in the experiment. For each *i-th* contact, levels were assigned for each element as $C_i=\{level(E_1), level(E_2), level(E_3), level(E_4), level(E_5), level(E_6)\}$. A serving system processed the assignment of visual intensity levels for elements. The structure of the object and intention to play was built by active elements with call-to-action messages and visual effects influencing users to play each game. Information about recommended games was represented in titles, thumbnails and extended form for featured games with a text description. The main factors related to the number of registered interactions were defined with $R^+$ as positive interactions represented by clicks and $R^-$ represented by negative responses and the number of attempts to remove interface.

**4.2 Assignment of users to test groups**

Different ways to display the object to measure the impact of visual intensity of the elements of interfaces on users' actions were implemented. Users were assigned to five groups during the first visit to the website; for each group, there was a different way of adjusting the levels of visual intensity selected. For the groups, the number of interactions representing response $R^+$ and $R^-$ were measured. With users assigned to groups, it was possible to observe how different ways of configuring the recommending interface affected the number of interactions. Users in the first three groups received the same object with each website reload without changing levels of visual intensity. Users in *group one*, the control group, received interfaces with elements on minimal level of visual intensity {1,1,1,1,1,1}. The interface design remained the same with each website refresh. Users in *group two* received a recommending object with a medium level of intensity configured as {3,3,3,3,3,3}. An interface with the highest level of intensity for each factor was configured as {5,5,5,5,5,5} for users in *group three*. For the next two groups, increasing and decreasing levels of intensity were assigned. In *group four*, levels of intensity increased with each refresh in 25 steps. The number of steps was limited to 25 as not many users had more than 25 contacts with the website and additional steps make data collection cumbersome. With first impressions to a user the object was generated with low levels 1:{1,1,1,1,1,1}. In a second impression, level of visual intensity was increased to 2:{2,1,1,1,1,1} and as follows for other impressions 3:{2,2,1,1,1,1}, 4:{2,2,2,1,1,1}, 5:{2,2,2,2,1,1}, 6:{2,2,2,2,2,1}, 7:{2,2,2,2,2,2}, 8:{3,2,2,2,2,2}, 9:{3,3,2,2,2,2}, 10: {3,3,3,2,2,2}, 11: {3,3,3,3,2,2}, 12:{3,3,3,3,3,2}, 13:{3,3,3,3,3,3}, 14:{4,3,3,3,3,3}, 15:{4,4,3,3,3,3}, 16:{4,4,4,3,3,3}, 17:{4,4, 4,4,3,3}, 18:{ 4,4,4,4,4,3}, 19:{4,4,4,4,4,4}, 20:{5,4,4,4,4,4}, 21:{5,5,4,4,4,4}, 22:{5,5,5,4,4,4}, 23:{5,5,5,5,4,4}, 24:{5,5,5,5,5,4}, 25:{5,5,5,5,5,5}. The levels of elements' visual intensity were left unchanged if a user reached 25 impressions. For *group five,* levels of visual intensity decreased with each website reload starting from highest 1:{5,5,5,5,5,5} and with next step 2:{4,5,5,5,5,5} and as follows for all other impressions 3:{4,4,5,5,5,5}, 4:{4,4,4,5,5,5}, 5:{4,4,4,4,5,5}, 6:{4,4,4,4,4,5}, 7:{4,4,4,4,4,4}, 8:{3,4,4,4,4,4}, 9:{3,3,4,4,4,4}, 10: {3,3,3, 4,4,4}, 11:{3,3,3,3,4,4}, 12:{3,3,3,3,3,4}, 13:{3,3,3,3,3,3}, 14:{2,3,3,3,3,3}, 15:{2,2,3,3,3,3}, 16:{2,2,2,3,3,3}, 17:{2,2,2,2,3,3}, 18 :{2,2,2,2,2,3}, 19:{2,2,2,2,2,2}, 20:{1,2,2,2,2,2}, 21:{1,1,2,2,2,2}, 22:{1,1,1,2,2,2}, 23:{1,1,1,1,2,2}, 24:{1,1,1,1,1,2}, 25:{1,1,1,1,1,1}.

## 5. Results of experiment based on changing levels of visual intensity

During the experiment, it was observed how changing the levels of visual intensity altered the relation between $R^+$ and $R^-$ response with results from the lowest visual intensity as a reference point. Analysis was performed for results obtained in groups of users with different interface configurations. Users made contact with the website content, and the settings of the interface were adjusted to the specific group. The first part of the analysis shows the results obtained with incremental intensity levels during user sessions. The second part of the study shows results obtained if the visual intensity of elements is decreased; in the last part, the results were compared for flat unchanged visual intensity.

### 5.1 Incremental levels of visual intensity

First analysed were the results from group $G_4$, where visual intensity was increased with each impression of the recommending interface and each time the user loaded the website leading up to 25 steps, from the lowest possible levels {1,1,1,1,1,1} for all elements up to {5,5,5,5,5,5} as described in the experiment design section. Aggregated results for the 25 contacts are shown in Table 1. Value one in column L represents results for first view with the lowest visual intensity, value two represents the second view and higher intensity, etc. A total of 399,146 views, shown in column V, were registered with 14,134 positive reactions $R^+$ represented by clicks within the interface and 983 negative reactions $R^-$. There were 106,329 page views registered in the system, resulting from the first contact with the interface. The second contact achieved 57,347 views, and the number of contacts dropped to 1,434 with C=25, representing users loading the interface 25 times, or the equivalent of visiting 25 subpages. Positive response factor $R^+F$ and negative response factor $R^-F$ for each contact and level of intensity of elements within the webpage were measured. Negative response rate $R^-R$ shows the relation of negative responses to positive responses. $CR^+$ and $CR^-$ factors represent rate changes for positive and negative responses, respectively, between contacts and are computed for each visual intensity level L using formula $CR^+(L) = R^+F(L)/R^+F(L-1)$ and $CR^-(L) = R^-F(L)/R^-F(L-1)$.

| L | $E_1$ | $E_2$ | $E_3$ | $E_4$ | $E_5$ | $E_6$ | V | $R^+$ | $R^-$ | $R^+F$ | $R^-F$ | $R^-R$ | $CR^+$ | $CR^-$ |
|---|---|---|---|---|---|---|---|---|---|---|---|---|---|---|
| 1  | 1 | 1 | 1 | 1 | 1 | 1 | 106329 | 2452 | 230 | 2.31% | 0.22% | 9.38%  | -      | -      |
| 2  | 2 | 1 | 1 | 1 | 1 | 1 | 57347  | 1736 | 159 | 3.03% | 0.28% | 9.16%  | 1.3117 | 1.2727 |
| 3  | 2 | 2 | 1 | 1 | 1 | 1 | 42070  | 1342 | 84  | 3.19% | 0.20% | 6.26%  | 1.0528 | 0.7143 |
| 4  | 2 | 2 | 2 | 1 | 1 | 1 | 32751  | 1321 | 99  | 4.03% | 0.30% | 7.49%  | 1.2633 | 1.5000 |
| 5  | 2 | 2 | 2 | 2 | 1 | 1 | 26135  | 1023 | 42  | 3.91% | 0.16% | 4.11%  | 0.9702 | 0.5333 |
| 6  | 2 | 2 | 2 | 2 | 2 | 1 | 21207  | 983  | 34  | 4.64% | 0.16% | 3.46%  | 1.1867 | 1.0000 |
| 7  | 2 | 2 | 2 | 2 | 2 | 2 | 17448  | 732  | 33  | 4.20% | 0.19% | 4.51%  | 0.9052 | 1.1875 |
| 8  | 3 | 2 | 2 | 2 | 2 | 2 | 14561  | 732  | 29  | 5.03% | 0.20% | 3.96%  | 1.1976 | 1.0526 |
| 9  | 3 | 3 | 2 | 2 | 2 | 2 | 12096  | 577  | 22  | 4.77% | 0.18% | 3.81%  | 0.9483 | 0.9000 |
| 10 | 3 | 3 | 3 | 2 | 2 | 2 | 10260  | 488  | 20  | 4.76% | 0.19% | 4.10%  | 0.9979 | 1.0556 |
| 11 | 3 | 3 | 3 | 3 | 2 | 2 | 8813   | 400  | 35  | 4.54% | 0.40% | 8.75%  | 0.9538 | 2.1053 |
| 12 | 3 | 3 | 3 | 3 | 3 | 2 | 7485   | 341  | 21  | 4.56% | 0.28% | 6.16%  | 1.0044 | 0.7000 |
| 13 | 3 | 3 | 3 | 3 | 3 | 3 | 6431   | 283  | 22  | 4.40% | 0.34% | 7.77%  | 0.9649 | 1.2143 |
| 14 | 4 | 3 | 3 | 3 | 3 | 3 | 5565   | 274  | 12  | 4.92% | 0.22% | 4.38%  | 1.1182 | 0.6471 |
| 15 | 4 | 4 | 3 | 3 | 3 | 3 | 4798   | 233  | 23  | 4.86% | 0.48% | 9.87%  | 0.9878 | 2.1818 |
| 16 | 4 | 4 | 4 | 3 | 3 | 3 | 4258   | 194  | 19  | 4.56% | 0.45% | 9.79%  | 0.9383 | 0.9375 |
| 17 | 4 | 4 | 4 | 4 | 3 | 3 | 3761   | 178  | 23  | 4.73% | 0.61% | 12.92% | 1.0373 | 1.3556 |
| 18 | 4 | 4 | 4 | 4 | 4 | 3 | 3261   | 152  | 13  | 4.66% | 0.40% | 8.55%  | 0.9852 | 0.6557 |
| 19 | 4 | 4 | 4 | 4 | 4 | 4 | 2917   | 145  | 10  | 4.97% | 0.34% | 6.90%  | 1.0665 | 0.8500 |
| 20 | 5 | 4 | 4 | 4 | 4 | 4 | 2541   | 123  | 12  | 4.84% | 0.47% | 9.76%  | 0.9738 | 1.3824 |
| 21 | 5 | 5 | 4 | 4 | 4 | 4 | 2291   | 114  | 12  | 4.98% | 0.52% | 10.53% | 1.0289 | 1.1064 |
| 22 | 5 | 5 | 5 | 4 | 4 | 4 | 2020   | 98   | 8   | 4.85% | 0.40% | 8.16%  | 0.9739 | 0.7692 |
| 23 | 5 | 5 | 5 | 5 | 4 | 4 | 1782   | 82   | 8   | 4.60% | 0.45% | 9.76%  | 0.9485 | 1.1250 |
| 24 | 5 | 5 | 5 | 5 | 5 | 4 | 1585   | 73   | 6   | 4.61% | 0.38% | 8.22%  | 1.0022 | 0.8444 |
| 25 | 5 | 5 | 5 | 5 | 5 | 5 | 1434   | 58   | 7   | 4.04% | 0.49% | 12.07% | 0.8764 | 1.2895 |

**Notes:** L- visual intensity level, $E_1$ - $E_6$ - visual intensities assigned to interface elements, V - number of interface views, $R^+$ - number of positive responses (clicks within interface), $R^-$ - number of negative responses (attempts to disable interface), $R^+F$ - Positive Response Factor as a percentage of positive interactions in relation to number of views, $R^-F$ - Negative Response Factor as a percentage of negative interactions in relation to number of views, $R^-R$ - Negative Response Rate showing number of negative responses in relation to number of positive responses, $CR^+$ - Rate Change between intensity levels for positive response, $CR^-$ - Rate Change between intensity levels for negative response.

**Table 1.** Statistics for group $G_4$ with increasing visual intensity of website elements

In Fig. 4, a chart representing growth of Positive Response Factor ($R^+F$) is presented. The maximal achieved point is at the level of five-percent, occurring at the tenth level of visual intensity. Further increases did not result in changes. Analysis of negative response $R^-F$ illustrated in Fig. 5 reveals growth through 25 levels with the lowest value equal to 0.16% and the highest value equal to 0.61%. Changes of visual intensity during the first nine steps did not result in growth of the R⁻F factor, but from the tenth level, growth was seen.

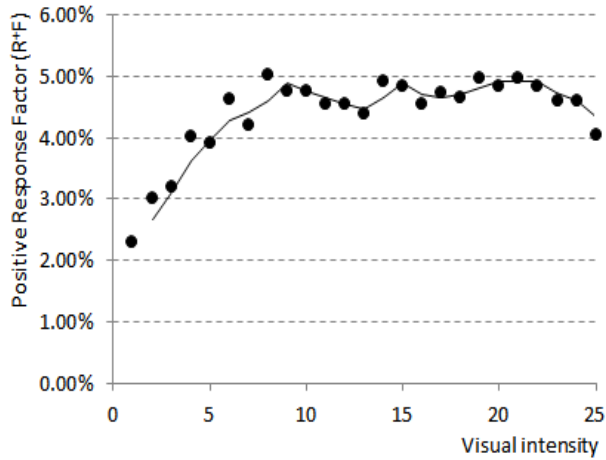 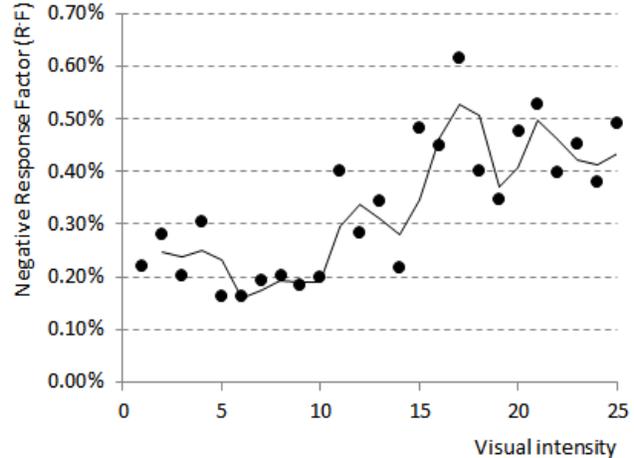

**Fig. 4.** Dynamics of changes of Positive Response Factor ($R^+F$) together with increasing level of influence.

**Fig. 5.** Dynamics of changes of Negative Response Factor ($R^-F$) together with increasing level of influence.

The mean value of the $CR^+$ factor is at the level of 1.0289 (SD=0.1110) for all contacts, while for first 10 contacts, it is maintained at a 1.0926 level (SD=0.1494). The mean value of changes is equal to 0.9907 (SD=0.0574) from the eleventh level, and no increase of positive responses is recorded. Analysis of changes rate for negative response $CR^-$ shows total mean value at the level of 1.0992 (SD=0.4130) and 1.0240 (SD=0.2893) for first 10 contacts and visual intensity levels, while the average value of negative response for levels of visual intensity greater than 10 shows 1.1443 (SD=0.4760) change factor. In the next step, the distance between normalised series with different levels of influence was computed using dynamic time warping method (Rabiner and Juang, 1993). Fig. 6 presents normalised responses showing regions with growth of positive and negative results. Computing differences between $R^+F$ and $R^-F$ resulted in detecting regions with the highest dynamic of changes and are presented in Fig. 7 based on differences computed for each level.

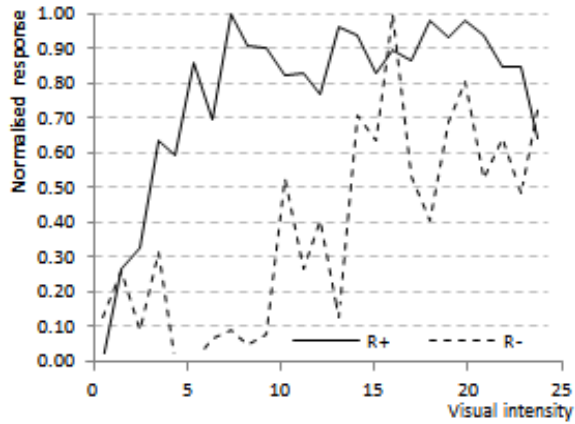 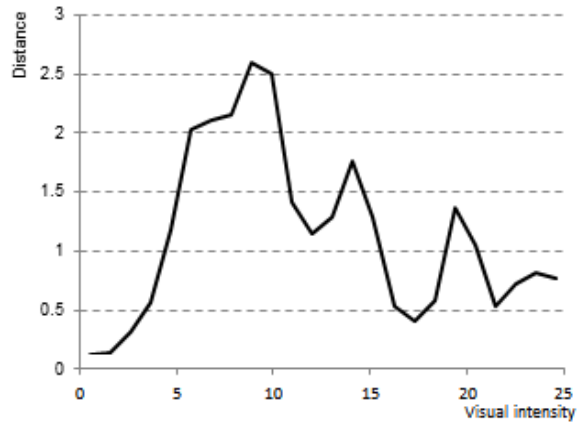

**Fig. 6.** Normalised response for incremental approach for Positive Response (R$^+$) and Negative Response (R$^-$)

**Fig. 7.** Distance between Positive Response (R$^+$) and Negative Response (R$^-$) for each level for increasing visual intensity

The total distance between measurement series was obtained at the level of 5.06. Distance was compared for different regions and for the segments from the first to the tenth level of influence, and distance was computed as 6.54; analysis for levels higher than 10 showed changes at the level of 5.59.

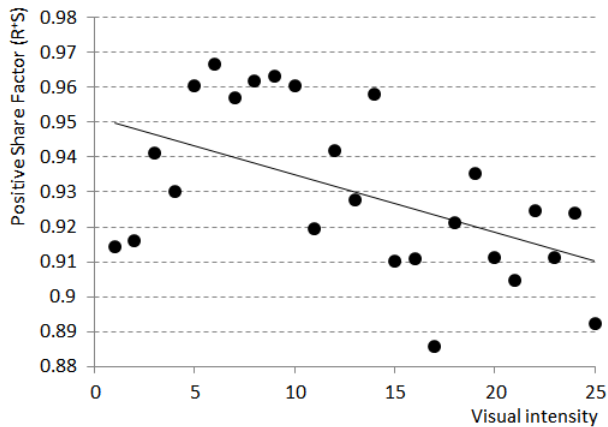 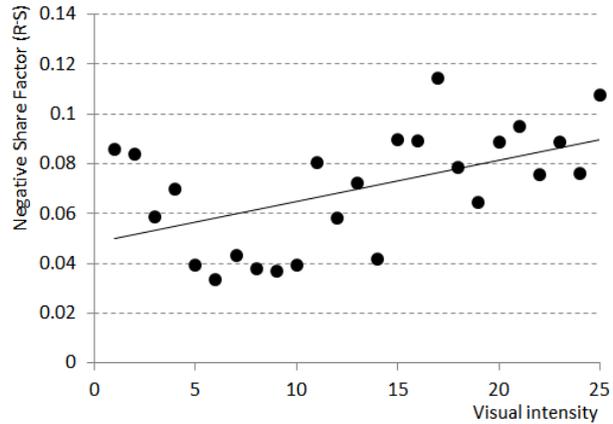

**Fig. 8.** Changes of Positive Share Factor (R$^+$S)

**Fig. 9.** Changes of Negative Share Factor (R$^-$S)

Additional analysis was performed to discern the relationship between the total number of interactions and positive/negative responses. Sharing factors such as R$^+$S and R$^-$S represent the share of positive and negative interactions divided by the total number of interactions. Fig. 8 shows that the share of positive interactions grows through the first ten levels of influence and then, after a brief stabilisation, the share drops. Another trend is observed for negative response share in Fig. 9, where negative share grows from level eleven. If influence is kept at the twelfth level, the results are at the accepted level, while the negative response is kept at the low levels.

**5.2 Decreasing levels of influence**

To perform an extended analysis, the results from group $G_5$ were used, with decreasing visual intensity with each contact. For group five, visual influence decreased from {5,5,5,5,5,5} to {1,1,1,1,1,1} in 25 steps; aggregated results are shown in Table 2, where L - represents the level of visual intensity, $E_1$-$E_6$ - elements within interactive component, V - number of views, $R^+$ - number of positive response (clicks in this case), $R^-$ - number of negative responses, $R^+F$ - positive response factor based on the number of clicks divided by the number of impressions, $R^-F$ - negative response factor based on number of negative responses divided by a number of impressions, $R^-R$ - negative response rate based on number of negative responses divided by a number of positive responses. $CR^+$ and $CR^-$ factors represent rate change for positive and negative responses, respectively, between contacts, and are computed for each visual intensity level L using formula $CR^+(L) = R^+F(L)/R^+F(L-1)$ and $CR^-(L) = R^-F(L)/R^-F(L-1)$.

| L | $E_1$ | $E_2$ | $E_3$ | $E_4$ | $E_5$ | $E_6$ | V | $R^+$ | $R^-$ | $R^+F$ | $R^-F$ | $R^-R$ | $CR^+$ | $CR^-$ |
|---|---|---|---|---|---|---|---|---|---|---|---|---|---|---|
| 1 | 5 | 5 | 5 | 5 | 5 | 5 | 95183 | 3009 | 939 | 3.16% | 0.99% | 31.21% | - | |
| 2 | 5 | 5 | 5 | 5 | 5 | 4 | 55945 | 2057 | 386 | 3.68% | 0.69% | 18.77% | 1.1646 | 0.6970 |
| 3 | 5 | 5 | 5 | 5 | 4 | 4 | 40982 | 1703 | 156 | 4.16% | 0.38% | 9.16% | 1.1304 | 0.5507 |
| 4 | 5 | 5 | 5 | 4 | 4 | 4 | 31709 | 1378 | 145 | 4.35% | 0.46% | 10.52% | 1.0457 | 1.2105 |
| 5 | 5 | 5 | 4 | 4 | 4 | 4 | 25379 | 1036 | 74 | 4.08% | 0.29% | 7.14% | 0.9379 | 0.6304 |
| 6 | 5 | 4 | 4 | 4 | 4 | 4 | 20454 | 882 | 76 | 4.31% | 0.37% | 8.62% | 1.0564 | 1.2759 |
| 7 | 4 | 4 | 4 | 4 | 4 | 4 | 16755 | 764 | 41 | 4.56% | 0.24% | 5.37% | 1.0580 | 0.6486 |
| 8 | 4 | 4 | 4 | 4 | 4 | 3 | 13899 | 586 | 40 | 4.22% | 0.29% | 6.83% | 0.9254 | 1.2083 |
| 9 | 4 | 4 | 4 | 4 | 3 | 3 | 11600 | 493 | 5 | 4.25% | 0.04% | 1.01% | 1.0071 | 0.1379 |
| 10 | 4 | 4 | 4 | 3 | 3 | 3 | 9735 | 414 | 11 | 4.25% | 0.11% | 2.66% | 1.0000 | 2.7500 |
| 11 | 4 | 4 | 3 | 3 | 3 | 3 | 8290 | 334 | 22 | 4.03% | 0.27% | 6.59% | 0.9482 | 2.4545 |
| 12 | 4 | 3 | 3 | 3 | 3 | 3 | 7102 | 305 | 5 | 4.29% | 0.07% | 1.64% | 1.0645 | 0.2593 |
| 13 | 3 | 3 | 3 | 3 | 3 | 3 | 6114 | 232 | 7 | 3.79% | 0.11% | 3.02% | 0.8834 | 1.5714 |
| 14 | 3 | 3 | 3 | 3 | 3 | 2 | 5294 | 193 | 1 | 3.65% | 0.02% | 0.52% | 0.9631 | 0.1818 |
| 15 | 3 | 3 | 3 | 3 | 2 | 2 | 4582 | 185 | 8 | 4.04% | 0.17% | 4.32% | 1.1068 | 8.5000 |
| 16 | 3 | 3 | 3 | 2 | 2 | 2 | 3973 | 138 | 7 | 3.47% | 0.18% | 5.07% | 0.8589 | 1.0588 |
| 17 | 3 | 3 | 2 | 2 | 2 | 2 | 3545 | 137 | 7 | 3.86% | 0.20% | 5.11% | 1.1124 | 1.1111 |
| 18 | 3 | 2 | 2 | 2 | 2 | 2 | 3096 | 112 | 6 | 3.62% | 0.19% | 5.36% | 0.9378 | 0.9500 |
| 19 | 2 | 2 | 2 | 2 | 2 | 2 | 2642 | 87 | 3 | 3.29% | 0.11% | 3.45% | 0.9088 | 0.5789 |
| 20 | 2 | 2 | 2 | 2 | 2 | 1 | 2313 | 68 | 2 | 2.94% | 0.09% | 2.94% | 0.8936 | 0.8182 |
| 21 | 2 | 2 | 2 | 2 | 1 | 1 | 2030 | 65 | 2 | 3.20% | 0.10% | 3.08% | 1.0884 | 1.1111 |
| 22 | 2 | 2 | 2 | 1 | 1 | 1 | 1837 | 62 | 3 | 3.38% | 0.16% | 4.84% | 1.0563 | 1.6000 |
| 23 | 2 | 2 | 1 | 1 | 1 | 1 | 1557 | 56 | 2 | 3.60% | 0.13% | 3.57% | 1.0651 | 0.8125 |
| 24 | 2 | 1 | 1 | 1 | 1 | 1 | 1387 | 33 | 1 | 2.38% | 0.07% | 3.03% | 0.6611 | 0.5385 |

| 25 | 1 | 1 | 1 | 1 | 1 | 1 | 1234 | 25 | 1 | 2.03% | 0.08% | 4.00% | 0.8529 | 1.1429 |

**Notes:** L- visual intensity level, $E_1$ - $E_6$ - visual intensities assigned to interface elements, V - number of interface views, $R^+$ - number of positive responses (clicks within interface), $R^-$ - number of negative responses (attempts to disable interface), $R^+F$ - Positive Response Factor as a percentage of positive interactions in relation to number of views, $R^-F$ - Negative Response Factor as a percentage of negative interactions in relation to number of views, $R^-R$ - Negative Response Rate showing number of negative responses in relation to number of positive responses, $CR^+$ - Rate Change between intensity levels for positive response, $CR^-$ - Rate Change between intensity levels for negative response.

**Table 2.** Total aggregated stats for group $G_5$ for all contacts

The interface with decreasing influence received total conversions at a rate of 3.70%, which was 84.18% of aggregated conversion from the increasing rate. The negative response factor at the level of 0.23% was 69.92% of value recorded in incremental approach. Figure 10 shows how decreasing levels of visual influence affected response rate. Changes were observed from 4.5% down to 2%. While the dynamic of dropping rates was relatively low, observed changes from the $R^-F$ factor in Fig. 11 show that during the first ten levels, the negative response dropped from 0.99% to 0.11%. Further decrease of influence did not result in a change in negative response. The results show that influence at about the twelfth level achieves an acceptable rate of interactions while keeping negative response at low levels.

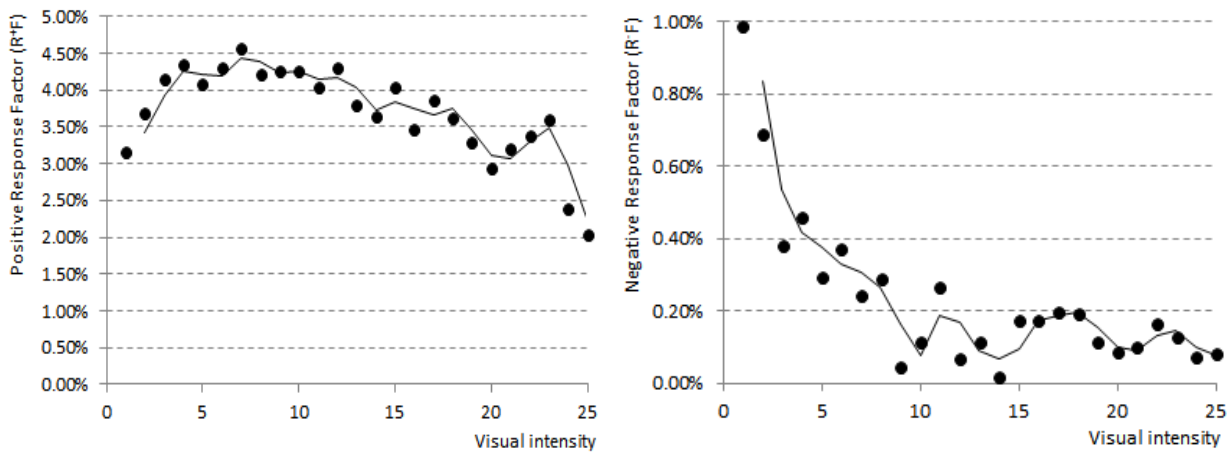

**Fig. 10.** Changes of Positive Response Factor ($R^+F$)  **Fig. 11.** Changes of Negative Response Factor ($R^-F$)

The mean value of the $CR^+$ factor is at the level of 0.9810 (SD=0.1103) for all contacts, while for first 10 contacts, mean value is maintained at 1.0121 level (SD=0.0675). After the tenth level, the mean value of changes is equal to 0.9609 (SD=0.1292) with no significant changes during the first ten levels. Analysis of changes rate for negative response $CR^-$ shows total mean value at the level of 1.3522 (SD=1.6836) and 1.2074 (SD=0.8775) for first ten contacts and visual intensity levels, while mean value of negative response for levels of visual intensity greater than ten, and shows a 1.4453 (SD=2.0734) change factor.

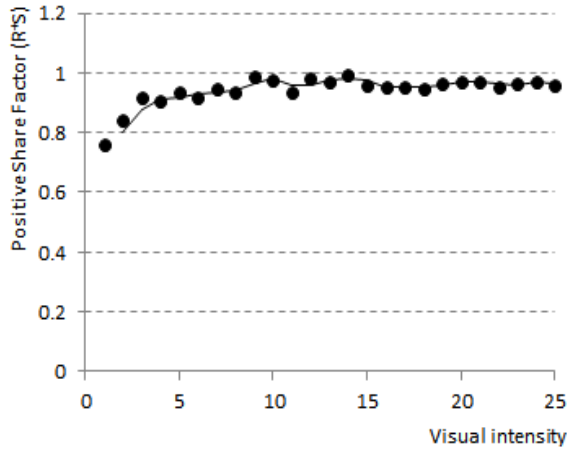 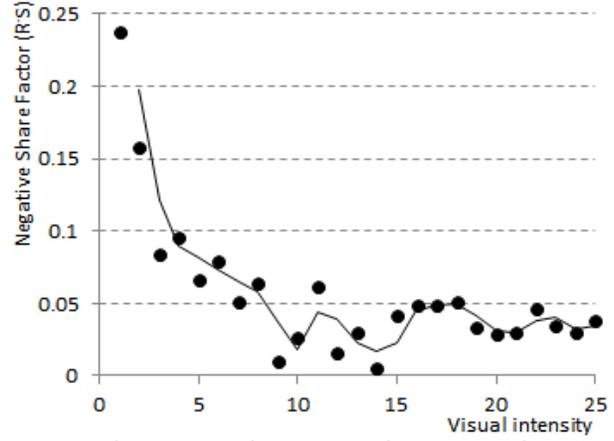

**Fig. 12.** Changes of Positive Share Factor ($R^+S$)   **Fig. 13.** Changes of Negative Share Factor ($R^-S$)

The analysis related to share of positive and negative relations shown in Figs. 12 and 13 reveals high dynamics from a dropping share of negative response during the first ten levels, while the growth of positive responses shows smaller dynamics. Fig. 14 presents normalised response for decreasing approach. Computing differences between $R^+F$ and $R^-F$ resulted in detecting regions with the highest dynamic of changes (Fig. 15) based on differences computed for each level.

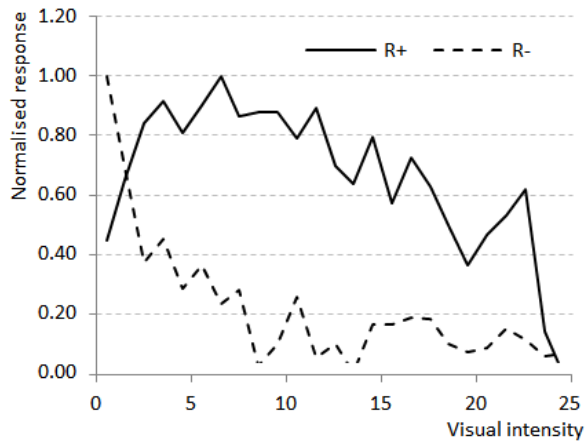 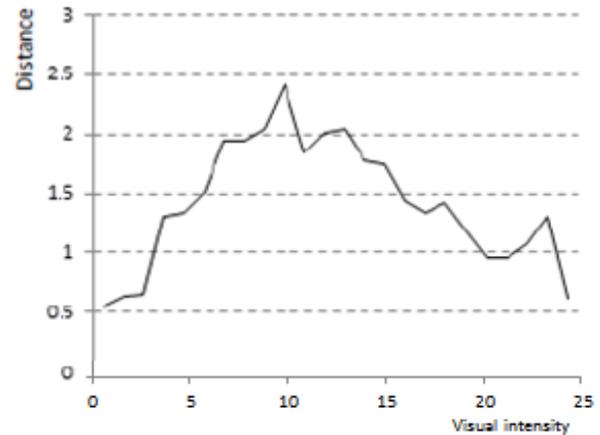

**Fig. 14.** Normalised response for decremental approach for Positive Response ($R^+$) and Negative

**Fig. 15.** Distance between Positive Response ($R^+$) and Negative Response ($R^-$) for each level of

Response (R⁻)                                      influence for decreasing intensity

Total distance between measurement series was computed at the level 4.78. The distance was compared for different regions, and for the segment from the first to tenth level of influence, distance was computed as 6.91. Analysis for levels higher than ten showed changes at the level of 6.99. The dynamic drop of negative response is accompanied by a slight drop of positive response $R^+$.

**5.3 Aggregated results for all approaches**

In groups $G_1$, $G_2$, and $G_3$, the object did not change, and the same settings were employed for users on low, medium and maximal levels, respectively, during their entire stay on the website. Table 3 shows aggregated results from all target groups including increasing and decreasing approaches.

| G | Visual intensity | V | $R^+$ | $R^-$ | $R^+F$ | $R^-F$ |
|---|---|---|---|---|---|---|
| $G_1$ | Min | 4141 | 105 | 9 | 2.54% | 0.22% |
| $G_2$ | Med | 4618 | 159 | 5 | 3.44% | 0.11% |
| $G_3$ | Max | 5920 | 243 | 41 | 4.10% | 0.69% |
| $G_4$ | Increasing | 399146 | 14134 | 983 | 3.54% | 0.25% |
| $G_5$ | Decreasing | 376637 | 14354 | 1950 | 3.81% | 0.52% |

**Notes:** G - target group, V - number of interface views, $R^+$ - number of positive responses (clicks within interface), $R^-$ - number of negative responses (attempts to disable interface), $R^+F$ - Positive Response Factor as a percentage of positive interactions in relation to number of views, $R^-F$ - Negative Response Factor as a percentage of negative interactions in relation to number of views.

**Table 3.** Aggregated statistics for groups $G_1$-$G_5$.

In group $G_1$, impressions were based on the lowest factors for all elements, and the levels of visual intensity did not change between page views. The means of presenting the recommending object was thought to be most neutral to users; this approach is represented in research related to content repetitions.

Test group $G_1$ delivered 4,141 impressions, and users played 105 recommended games. The system registered nine negative actions represented by attempts to remove the interface in this group. The obtained metrics were $R^+F(G_1)$=2.54%, $R^-F(G_1)$=0.22%. In $G_2$, with medium visual intensity the number of impressions being 4,618, the recommending object received 159 clicks and five negative responses.

Conversion $R^+F(G_2)$ went to 3.44%, which is 35% higher than conversion in $G_1$. The negative response rate was at level 0.11%, even lower than for group $G_1$. In $G_3$, with maximal influence on users and all impressions configured with {5,5,5,5,5,5}, a higher conversion $R^+F$ ($G_3$) was observed, and the rate increased to 4.1%, but as a side effect, the negative response rate went up to 0.69%. Comparing the results for $G_3$ with those of $G_2$, there is a 19.19% increase in $R^+F$. $R^-F$ factor is 6.27 times of $G_2$ and 3.14 times of $G_1$ respectively. A slight increase of conversion was connected with a high increase of $R^-F$, and as a result, there was a decline in user experience. Decreasing approach for group $G_5$ resulted in slightly better results than for group $G_4$ with increasing approach. The negative response factor $R^-F$ for $G_5$ was two times higher than for group $G_4$.

## 6. Discussion

This study explored how gradual changes of visual intensity affected users' interactions within the interface. It shows that it is possible to increase users' interaction with the interface by adjusting visual intensity of the interface elements. However, interaction can only be increased towards a point where visual intensity becomes detrimental. Therefore, our study was set out to find an optimal point of visual intensity where the users' interaction would be maximised. Adding attention-catching elements such as high-frequency flashing components resulted in increasing negative reactions. Further increasing the visual intensity of the object results in a higher number of negative reactions versus positive effects. The results showed that a higher dynamic growth was observed for negative reactions than for positive reactions. The results showed the changes in the number of interactions between levels of visual intensity. Negative response grew faster than conversion after increasing visual intensity of the web interface. A saturation point, with coexistence between maximum conversion and minimum negative response, can be found.

The proposed approach for constructing website elements with a call-to-action purpose can be used in the analytical process of searching for the optimal design. Therefore, increasing the visual intensity to increase the number of interactions should be done carefully since too high visual intensity has proportionately larger effect on negative reactions. Careful efforts to attract user attention within webpages should be made to find a site-specific sweet spot by incrementally adjusting visual intensity to reduce adverse influence on user experience.

The research delivered theoretical and practical implications and supported assumptions that a multi-dimensional increase of visual intensity levels caused by the implementation of flashing with high-frequency elements in the message does not necessarily achieve better results. The compromise result can be based on the selection of design variants delivering user experience and effects at an acceptable level. Excessive exploitation of a website can lead to lower user comfort levels, loss of audience, and failure to generate enough positive results to compensate for adverse side effects. An approach that enables determination of optimal visual intensity levels and the scope to which it reflects the acquired results is justified. This research shows measures and representation of both positive and negative response. The performed research provides a foundation and tools for further theoretical work and experimentation.

**7. Conclusions**

This study introduced a method for adjusting and finding an optimal point of visual intensity of website elements based on a gradual approach. The proposed method enabled analysis of specific interface elements and their impact on consumer conversion. It helped determine the level of visual intensity of website elements through incremental and decremental adjustments and tracked the changes in user behaviour.

Apart from previously mentioned theoretical aspects, the solutions offer several practical contributions. Website operators and marketers can use them to search for optimal visual intensity of interface elements. Preparing content considering gradual influence can limit the negative impact on users while maintaining final performance. The study showed that moderate levels of visual intensity could result in an acceptable level of conversions and improvements are observed till seventh from twenty-five levels of intensity. Further increases in visual intensity result in a negative response and decreasing user experience, while conversion improvement was not observed.

The final scope of practical applications is dependent on the range of variation with monitored elements and designer expectations. Introducing the decomposition of the object into its constituent parts makes it possible to monitor the effects on the level of invasiveness of interactive elements. An adopted level can be set in the decision-making process and is dependent on designers' preferences towards sustainable marketing solutions.

The gradual approach does have limitations. First, it requires at least several events, such as page views in sequence, to deliver variants with varying intensity. Users with low sequence lengths will be presented low intensity content without a chance to display more intensive messages, resulting in possible low conversion. The field experiment conducted also has practical limitations, as it used only scaling of visual intensity of elements within the website. Apart from visual intensity, other parameters could be utilised including the size of marketing content, localisation within the website.

Proposed approach offers several directions for the future work, which could target extending the research to other forms of influence on web users based on persuasion or intrusive components. It could be used for audio or video content intensity as well. Automated detection of acceptable levels of influence and verification of scenarios other than increasing and decreasing influence, e.g., pulse scenario, can be considered. Another direction for future research includes measuring influence differences between levels and adjusting them according to obtained results.

**Acknowledgments**

The work was supported by the National Science Centre of Poland, the decision no. 2017/27/B/HS4/01216